\documentstyle[preprint,eqsecnum,aps]{revtex}

\begin{document}
\title{ Umklapp Processes for Electrons and Their Renormalisation
Group Flow}
\author{ Deepak Kumar and   R . Rajaraman\cite{byline}}
\address{ School of Physical Sciences\\
 Jawaharlal Nehru University\\
 New Delhi,  INDIA 110067}
\maketitle\
\begin{abstract}
We study Umklapp couplings and their renormalisation group flow
for electrons in a two-dimensional lattice. It is shown that the
effective low energy  Hamiltonian involves not only
forward scattering, but also considerable non-forward Umklapp
scattering, when the  electrons fill a  substantial fraction of
the valence band.  The behavior of these couplings under the
renormalisation group is  studied at the tree and one-loop
level. It is shown that they remain marginal. We
conclude  with the possible consequences of this on
properties of such electronic systems and their Fermi liquid
theory.
\end{abstract}
\pacs{05.30 Fk , 71.10.+x}

\section{ Introduction.}
In recent years the renormalisation group (RG) method has been
used to give a simple microscopic justification of the
postulates of the Landau theory of Fermi liquids
\cite{Shank1,Shank2,Polch,Ben,FeldTru,FeldMag}. In particular it
has been shown that for purposes of the low-energy excitations
near the Fermi surface, the  fermion interaction can be
characterised by  just a few relevant couplings. For
\underline{normal} Fermi liquids in two dimensions  without BCS
interactions, it has been argued that, the only   interaction
that is marginal corresponds to forward scattering
and its exchange.  Roughly speaking , this happens because (a)
low-energy physics involves only states lying arbitrarily close
to the Fermi surface and (b) for such states on the Fermi
surface momentum conservation permits only forward scattering
and its exchange.  Careful RG analysis supports this conclusion
as shown in Shankar's lucid review of this topic \cite{Shank1}.
It is this forward scattering coupling constant that forms the
microscopic basis for Landau's phenemenological Fermi liquid
parameter in 2 dimensions \cite{Landau,AGD,Noz}.

While this argument is good at sufficiently low densities, at
higher densities when the Fermi surface approaches the Brilluoin
zone boundary, additional lattice effects come into play. These
are related to Umklapp processes, where momentum is conserved
only modulo reciprocal lattice vectors.  In such a process, a
pair of particles lying on or near the Fermi surface can scatter
in specific non-forward directions, in addition to the usual
forward scattering \cite{AAA}.

 In this paper we will analyse the conditions under which such
Umklapp couplings arise to a significant extent and study the RG
flow of such couplings. We will see that circumstances where
substantial Umklapp scattering takes place are by no means
extraordinary. For typical systems, Umklapp processes will shown
to be permitted for some electrons on or near the Fermi surface
already at densities well below half-filling. Around
half-filling, almost all  pairs of states on the Fermi surface
will be permitted some Umklapp scattering, in addition to
forward scattering. Systems with such filling factors in the
valence band are not uncommon. For example in Alkali metals,
which are nice examples of one band free electron-like Fermi
surfaces, the ratio of Fermi surface radius to the distance of
the zone face closest to the origin is 0.88 \cite{AM}.

We will also study the flow of these Umklapp couplings  under
the renormalisation group process, both at the tree level and
the one-loop level.  We will see that at the tree level they are
marginal under the RG flow and thus survive. But they
do not make any further contribution at the one-loop level to
either their own RG flow equation or to the flow equation of
either the forward or BCS couplings.

This paper is organised as follows: In the next section we will
set up the preliminaries and study the RG flow of the couplings
at the tree level.  For simplicity, we will work in two
space-dimensions although it will be clear that our analysis and
conclusions can be generalised to 3-dimensions as well. To be
specific we will work with a square lattice and since spin is
not an essential complication to our arguments, we will study
spinless one-band Hamiltonians, just as was done in ref 1.  The
kinetic and the quartic interaction terms in the starting
Hamiltonian will be taken to be quite general. The Fermi surface
need not be isotropic ( circular) ; it is sufficient for our
analysis if it is convex and time-reversal invariant -- which
includes a  fairly large class of systems. Following this we do
the RG analysis at the tree level and show that even in the
presence of Umklapp, the only relevant couplings are those that
satisfy (lattice) momentum conservation on the Fermi surface.
 
Armed with this result, in sec.III, we will discuss the
kinematics of Umklapp processes, for particles lying
\underline{on} the Fermi surface. A pictorial criterion will
 be given of the fraction of initial states that can suffer
Umklapp scattering.  This will establish our basic claim that
significant Umlapp can take place at densities near
half-filling. In sec. IV, we will study the one-loop RG flow of
Umklapp couplings. It will be shown that while they remain
marginal at the tree level, at one-loop level, they
do not to contribute to either their own flow equation or those
of the forward or BCS couplings. This is followed by a
discussion on the physical implications of our result in Sec.V.

\section{The Model and Tree Level RG } The aim of this section
is to obtain conditions , upto tree level, on 4-point Umklapp
couplings so that they may remain marginal under RG
flow. The analysis in this section closely follows that of
Shankar \cite{Shank1} , adapted to handle anisotropic Fermi
surfaces and Umklapp processes.  We will omit  intermediate
steps but the major steps will be shown, with all the
definitions needed for completeness.  We begin with a fairly
general one-band system of spinless electrons in a two
dimensional lattice with chemical potential $\mu$ and a
Hamiltonian given by
\begin{eqnarray} {\cal H} \ \ &=& \ \
\sum_{{\vec K} \epsilon S_{\Lambda}} \ ( E({\vec K})- \mu)
 C_{\vec K}^{\dag} C_{\vec K}
  \ + \ \ { 1 \over N} \sum_{{\vec K}_1,{\vec K}_2,{\vec K}_3, {\vec
K}_4 \epsilon S_{\Lambda}} U({\vec K}_1,{\vec K}_2,{\vec K}_3, {\vec
K}_4) \nonumber \\
 & & \times C^{\dag}_{{\vec K}_4} C^{\dag}_{{\vec K}_3} C_{{\vec
K}_2} C_{{\vec K}_1}
\big[ \sum_{i} \delta ( { \vec K}_4 + {\vec K}_3 - {\vec K}_2 -
{\vec K}_1  - {\vec G}_i) \ \big] \label{H} \end{eqnarray}
 where the notation is as follows:\\
$C^{\dag}_{\vec K} $ and $C_{\vec K}$ denote the creation and
annhilation operators for the single particle eigenstates of
wavevector $\vec K$ . Though in principle the wavevector ${\vec
K}$  ranges over the entire first Brillouin zone of the lattice,
we will be  interested only in the low energy excitations of the
system near the Fermi surface $E({\vec K}) = \mu$. Therefore, as
is usually done the ${\vec K}$-sums are restricted in the model
to an energy shell $S_{\Lambda}$ defined by $\mid E(\vec K) -
\mu \mid \  \leq \ \Lambda$. This is what is meant by
$S_{\Lambda} $ in the sums in eq.(\ref{H}). 

 Given these energy shell conditions and the fact that we  want
to consider non-isotropic $E({\vec K})$ as well, we will
sometimes find it convenient to change variables on the ${\vec
K}$-plane from ${\vec K} \ = \ (K_{x}, K_{y})$ \ to $(\epsilon,
\theta)$ where $\epsilon ({\vec K}) \equiv E({\vec K})-\mu$, and
$\theta \equiv tan^{-1} {K_{y} \over K_{x}}$, with $J(\epsilon,
\theta)$ as the Jacobian of this tranformation.  ( We will
ignore possible Van Hove singlarities \cite{AM} in the Jacobian
J as well as nesting effects since these are not directly
related to our primary interest on Umklapp processes. Those
effects will have to be handled in addition  if needed.) The
function $U({\vec K}_1 ,{\vec K}_2 ,{\vec K}_3 , {\vec K}_4 ) $
denotes the 4-point interaction coupling  in the single particle
wavevector basis.  In the second term of Eq.(1), momentum
conservation $\delta$-functions are inserted as appropriate for
the periodic lattice, i.e.  where ${\vec G}_i$`s are all
possible reciprocal lattice vectors,
\underline{including the null vector}.  When ${\vec G}_i =  0$,
we have the usual normal process, but Umklapp processes with
${\vec G}_i \neq 0$, are also included wherever allowed. In fact
our main interest in this paper lies in such Umklapp
processes. We place no restriction on the form of $E({\vec
K})$ other than that it ,\\ (i) be time-reversal invariant, i.e.
\begin{equation}   E({\vec K}) \ \ = \ \ E(-{\vec K}),
\end{equation} and, (ii) give rise to a Fermi surface [FS]
defined by $E({\vec K}) = \mu$ which is \underbar{convex}.

       The renormalisation group [RG] procedure has been
described in ref.\cite{Shank1} in detail. For completeness we
will outline the major steps and concentrate on what it does to
Umklapp couplings.  The zero temperature grand canonical
partition function Z is first written as a path integral over
Grassmann variables. One has,
\begin{equation} Z \ \ = \ \ \int \ \bigg(\prod_{| \epsilon|
 < \Lambda}
d {\bar\psi}(\omega , \epsilon, \theta)  \  \ 
d \psi(\omega,\epsilon,\theta) \bigg)  \  \ exp(-(S_{0}+S_{I}))   
\end{equation}
where,
\begin{equation} S_{0} \ = \ \ \int_{-\Lambda}^{\Lambda}
 {d \epsilon  \over 2\pi}
\int_{-\infty}^{\infty} { d\omega \over 2\pi} \ \int_{0}^{2\pi} 
{d\theta \over 2\pi}
 \ \  {\bar\psi}(\omega , \epsilon, \theta) \  \big(i\omega-
 \epsilon \big) 
 \  \psi(\omega,\epsilon , \theta) \  J( \epsilon, \theta) 
 \label{S0} \end{equation}
\begin{eqnarray} S_I \ \ &=& \ \ {1 \over 4} \sum_{i} \bigg
[\prod_{j=1}^{3} 
\int_{-\Lambda}^{\Lambda} {d \epsilon_{j} \over 2\pi}
\int_{-\infty}^{\infty} { d\omega_{j} \over 2\pi} \
\int_{0}^{2\pi} { d\theta_{j} \over 2\pi} \ \ J(\epsilon_{j},
\theta_{j}) \ U({\vec K}_{1},{\vec K}_{2},{\vec K}_{3},
 {\vec K}_{4i}) \nonumber\\
 & & \times  {\bar\psi}(\omega_{4} , \epsilon_{4i}, \theta_{4})
  \ {\bar\psi}(\omega_{3} , \epsilon_{3}, \theta_{3})
  \  \psi(\omega_{2} , \epsilon_{2}, \theta_{2}) \ 
 \psi(\omega_{1} , \epsilon_{1},\theta_{1}) \ 
 \Theta (\Lambda \ - \ |\epsilon _{4i}|) 
  \bigg] \label{SI} \end{eqnarray}
In eq(\ref{SI}), all momenta  $ {\vec K}_{j}$ are  
$ {\vec K}_{j} (\epsilon_{j}, \theta_{j})$.  \ 
   $\omega_{4}$ and $ {\vec K}_{4i}$ are
understood to be fixed by conservation conditions
 \begin{equation} \omega_{4} \ = \ \omega_{1} + \omega_{2}
  - \omega_{3} \end{equation} and
\begin{equation} { \vec K}_{4i}  \ = \  {\vec K}_1 +{\vec K}_2
  + {\vec G}_i - {\vec K}_3 \label{Kcons} \end{equation}
 for each reciprocal vector ${\vec G}_{i}$.\\
The requirement that particle 4 also lie in the energy shell 
 $S_{\Lambda}$ has been implemented  in (\ref{SI}), following ref.\cite 
{Shank1}  by the step-function $ \Theta (\Lambda \ - \ |\epsilon_{4i}|)$.

Clearly , with ${\vec K}_{4i}$ already fixed by (\ref{Kcons}),
 this shell condition
on particle 4 in fact becomes a condition on particle 3.
 For a given ${\vec G}_{i}, \  {\vec K}_{1}$
and $ {\vec K}_{2}$ , some or no values of ${\vec K}_{3}$ may 
exist which place
${\vec K}_{4i}$ within the shell $S_{\Lambda}$.

The steps of the RG procedure are to first integrate out field
variables in the subshells between $S_{\Lambda}$ and $S_{\Lambda
/ s}$, where $s>1$ is the scaling parameter. Then we use
rescaled variables $\omega '=s\omega, \epsilon ' =s\epsilon$, and
\ $\psi ' (\omega ',\epsilon ', \theta) \ = \ (s)^{{-3 \over 2}}
\psi  (\omega ,\epsilon , \theta)$, so that the shell thickness
is restored to $\Lambda$ and the kinetic term ($\bar{\psi}
i\omega \psi$) in $S_0$ is left invariant. Under this rescaling
of energy variables, the corresponding rescaled momenta will be
$  {\vec K'}_{j} \ \equiv {\vec K}_{j} (\epsilon_{j}',
\theta_{j})$.  One also replaces the shell condition by an
exponential damping factor $ exp \ {-|\epsilon_{4i}| \over
\Lambda}$. See ref.\cite{Shank1} for a full description.

Under these RG transformations, in terms of rescaled variables,
 the shell condition returns to $S_{\Lambda}$. The renormalised
 quartic interaction becomes  
at the tree level, 
\begin{eqnarray} S'_I \ \ &=& \ \ {1 \over 4} \sum_{i} \bigg[
\prod_{j=1}^{3} 
\bigg(\int_{-\Lambda}^{\Lambda} {d \epsilon_{j}' \over 2\pi}
\int_{-\infty}^{\infty} { d\omega_{j}' \over 2\pi} \
\int_{0}^{2\pi} { d\theta_{j}' \over 2\pi} \ \ J\bigg( {\epsilon '
 \over s},   \theta \bigg)\bigg) U'({\vec K}_{1}',{\vec
K}_{2}',{\vec K}_{3}', {\vec K}_{4i}') \nonumber\\ & & \times
{\bar\psi}'(\omega_{4}' , \epsilon_{4i}', \theta_{4}) \
{\bar\psi}'(\omega_{3}' , \epsilon_{3}', \theta_{3}) \
\psi'(\omega_{2}' , \epsilon_{2}', \theta_{2}) \ \psi'(\omega_{1}'
, \epsilon_{1}',\theta_{1}) \ exp \ (-{|\epsilon_{4i}'| \over
\Lambda}) \bigg] \label{SI'} \end{eqnarray}
 where  the renormalised coupling $U'$ is given by,
\begin{equation} U'({\vec K}_{1}',{\vec K}_{2}',{\vec K}_{3}',
{\vec K}_{4i}') \ \equiv \  U({\vec K}_{1},{\vec K}_{2},{\vec
K}_{3}, {\vec K}_{4i})  exp \ \bigg( {-(s-1) | \epsilon _{4i}|
\over \Lambda} \bigg) \label{U'}\end{equation} The following
results hold in the $s \rightarrow \infty$ limit : (i) Only the
$\epsilon '$-independent term in $ J( {\epsilon ' \over s},
\theta)$ will remain marginal allowing us to
replace $J( {\epsilon ' \over s}, \theta)$ by $J(\theta)$ in
subsequent discussion. \\
 (ii) More importantly, in (\ref{U'}), the survival of the
renormalised coupling $U'$ requires in the $s \rightarrow
\infty$ limit that $\epsilon_{4i}=0$.  Recall that
$\epsilon_{4i}$ depends on ${\vec K}_{4i}$ which in turn is
determined by lattice momentum conservation (\ref{Kcons}).
Hence $\epsilon_{4i} =0$ forces a condition on ${\vec K}_{3}$
for a given ${\vec K}_{1}$ ${\vec K}_{2}$ and ${\vec G}_{i}$. To
unravel this condition on ${\vec K}_{3}$ it is useful to define
\begin{eqnarray} {\vec k} & \equiv  \  \ {\vec K} \ - \ {\vec
K}_{F}(\theta) \nonumber \\ \epsilon({\vec K}) \ &= \ E({\vec
K})- \mu \ \equiv v_{F}(\theta) \  k \end{eqnarray} for each
${\vec K}$, where ${\vec K}_{F}(\theta)$ is a vector
\underline{on} the Fermi surface in the same direction $\theta$
as ${\vec K}$, and $v_{F}(\theta)$ is the Fermi velocity at that
angle. Under rescaling we have $\epsilon'= s\epsilon, \ k' \
=sk$.  Given all this,
\begin{eqnarray} |\epsilon_{4i}| \ &=& \  v_{F}(\theta_{4i})
 |K_{4i} - K_{F}(\theta_{4i})| \nonumber \\
 &=&  \ v_{F}(\theta_{4i}) \big| |{\vec Q_{i}} +
{\vec q}| - K_{F} (\theta_{4i})  \ \big|\nonumber \\
&=& \ \ 0 \ \label{ep4i} \  \end{eqnarray}
 where 
 \begin{eqnarray} \vec {Q_{i}} &=& {\vec K}_{F}(\theta_{1}) +
{\vec K}_F(\theta_2)
 - {\vec K}_{F}(\theta_3) + \vec {G}_i \nonumber \\
 {\vec q} \ &=& \ {\vec k}_{1} + {\vec k}_{2} - {\vec k}_{3}
  \end{eqnarray}
Since $q=q'/s $ ,  (\ref {ep4i})
 reduces to \begin{equation} |Q_{i} \ - \ K_{F}(\theta_{4})| \ 
 = \ O(1 / s)  \end{equation}
In the $s \rightarrow \infty $ limit, this gives
\begin{equation} {\vec K}_F(\theta_{4i}) = {\vec K}_F(\theta_1)
 + {\vec K}_F(\theta_2) - {\vec K}_F(\theta_3) + \vec {G}_i 
 \label{FSeq} \end{equation} 
Note that this equation looks like a (lattice) momentum
conservation equation for any given reciprocal lattice vector
${\vec G}_i$. Of course, $ {\vec K}_F(\theta_{j}) $ is not the
same vector as ${\vec K}_{j} $ , but it has the same direction
and lies on the Fermi surface. Thus eq.(\ref{FSeq}) is to be
viewed as a condition on the final angles $\theta_3$ and
$\theta_4$ given the initial angles $\theta_1$ and $\theta_2$
and some  ${\vec G}_i$.

 As far as the dependence on energy is concerned, once
(\ref{ep4i}) is satisfied, the exponential factor in (\ref{U'})
becomes unity, and the remaining expresion for the renormalised
coupling $U'$ obeys just,
 \begin{equation} U'\big(\{ {\vec K}_j'
\} \big) \ = \  U\big(\{ {\vec K}_j \} \big)\end{equation} which
when    written as a function of $(\epsilon, \theta )$
variables, becomes
\begin{eqnarray} U'\big( \{ \epsilon_j , \theta_j \} )
&= \  U \big( \{ \epsilon_j , \theta_j \} \big) \nonumber \\
&= \  U \big( \{ \epsilon_{j}' / s, \theta_j \} \big) 
\end{eqnarray}
When expanded in powers of 1/s, this shows that in the RG $(s
\rightarrow \infty)$ the energy-dependent parts of $U'$
 becomes irrelevant, and $U' \ = \ U'\big
( \theta_1 , \theta_2 ,\theta_3 ,\theta_4  \big)$. All one
 needs to do is to find the allowed dependence of $U'$ on the
angles, using (\ref {FSeq}).

Such results had already been obtained in \cite{Shank1} for the
isotropic non-Umklapp case. There it was shown that only forward
scattering (and its exchange) remain marginal for
normal liquids. The forward coupling will then depend only on
the initial angles and was denoted in ref.[1] by the function
$F(\theta_1 ,\theta_2)$.  What we have done so far is only to
ensure , for completeness, that a similar on-Fermi-surface
condition on the angles obtains  more generally.  Our new
results on Umklapp couplings come starting with the next section
where we  find the allowed values (if any) of $\theta_3$ and
$\theta_4$ when ${\vec G}_i \neq 0$, by solving (\ref{FSeq})
\underline{on} the Fermi surface.

\section{ Kinematics of Umklapp Processes.}

In this section let us explore the angle condition (\ref{FSeq}).
We recall that this is just a lattice momentum conservation
condition on four momentum vectors  \underline{on} the Fermi
surface for any given lattice vector ${\vec G}_i $. To be
specific, let us take a square lattice. Its reciprocal lattice
vectors are $G_{\pm 1} \ = \ (\pm 2\pi,0)$ and $G_{\pm 2} \ = \
(0, \pm 2\pi)$  in units of inverse lattice spacing.  In 2-D,
the Fermi surface can be taken to be a closed curve, described
in polar coordinates by $ r \ = \ f(\theta)$, with $f(\theta) \
= \ f(\theta \ + \ \pi))$ by virtue of the time-reversal
condition (2.2).

	To start with, let us prove a small Lemma about the
domain of the total momentum of a pair of particles on the Fermi
surface.

\underbar {Lemma}

	Consider a pair of particles with momenta $ {\vec
K}_{1}$ and ${\vec K}_{2}$ lying \underbar{on} the Fermi surface
$r \ = \ f(\theta)$, corresponding to vectors OC and OD in
figure 1a. Let their total momentum $ {\vec P} \  = \ \ {\vec
K}_{1} \ + \ \ {\vec K}_{2} $ lie at some angle $\theta_{P}$
with magnitude $ \mid \vec{P} \mid \ = \ $ OP  as shown. Then by
the assumed convexity,
\begin{equation} OP \ = 2 (OB)  \ \leq \ 2f(\theta_{P})
 \end{equation}
Hence the sum of any two momenta on the Fermi surface always
lies in the area bounded by the curve $r \  =  \  2f(\theta)$.
Conversely, consider any vector ${\vec P}$ lying on or inside
the curve $r \ = \ 2f(\theta)$, and denoted by OP in fig 1b.
Draw another Fermi surface centered at P. Since $ OP  \ \leq \
2f(\theta_P) \ = \ 2(OL) \ = \ OL + PA$, these two Fermi
surfaces will generically intersect at two points C and D.  By
time-reversal symmetry, the vectors OD and CP are equal.  Hence
${\vec P}$ will be the sum of two momenta $ {\vec K}_{1}$ and
${\vec K}_{2}$  on the original Fermi surface denoted by OC and
OD respectively. Altogether, we see that the domain of the sum
of all possible pairs of  vectors on the Fermi surface is the
entire area bounded by $r \ = \ 2f(\theta )$.  For convenience
let us henceforth call this curve $r \ = \ 2f(\theta )$ the
"double Fermi surface" (DFS) and the area enclosed by it as the
"double Fermi volume" (DFV).

 [Although our discussion will mostly deal with 2 dimensional
systems, it is useful to note that a similar statement about the
domain of ${\vec P}$ can also be made for a 3 dimensional system
with a convex time-reversal invariant Fermi surface $r \ = \
f(\theta, \phi)$, using the same arguments.  The total momentum
of any pair of particles on the Fermi surface will once again
lie in the DFV bounded by the DFS \  $r \ = \ 2f(\theta, \phi)$.
Conversely, any ${\vec P}$ in this volume will be a sum of some
pair of momenta on the Fermi surface. In fact in 3-D, the
analogue of fig 1b will have two Fermi surfaces intersecting on
a whole curve and not just 2 points, so that any given ${\vec
P}$ will correspond to a whole family of pairs of momenta $
{\vec K}_{1}$ and ${\vec K}_{2}$  on the original Fermi surface
.]
 
Returning to 2 dimensions, it will be helpful to first recall
the familiar case of non-umklapp processess, with all momenta
lying on the Fermi surface. In a non-umklapp scattering process
momentum conservation requires
\begin{equation}  {\vec K}_{1} \ + \ {\vec K}_{2} \ \ =
\ \ {\vec P} \ \ = \ {\vec K}_{3} \ + \  {\vec K}_{4} 
\end{equation}
  In the absence of nesting, a generic non-zero total momentum
${\vec P}$  will be the sum of a unique pair of momenta on the
Fermi surface, such as the pair $ {\vec K}_{1}$ and ${\vec
K}_{2}$ in figures 1a and 1b.  Hence, the final pair of momenta
have to be the same as the initial pair. Therefore, only forward
or its exchange (backward) scattering is allowed on the Fermi
surface. The corresponding tree level renormalised coupling $U'$
is some function $F(\theta_{1}, \theta_{2})$ where recall that
$\theta_{i}$ refers to the polar angle of the ${\vec K}_{i}$.
The exception to this rule is the case $ {\vec P} \ = \ 0$
Then the two Fermi surfaces in Fig 1b will coalesce, and the
final momenta could be any pair of equal and opposite vectors on
the Fermi surface. This corresponds to the "BCS" coupling
denoted by $V(\theta_{1}, \theta_{3})$. In the absence of
Umklapp, these two functions   $F(\theta_{1}, \theta_{2})$ and
$V(\theta_{1}, \theta_{3})$ are the only tree-level four-fermion
couplings that remain marginal as has been shown in
\cite{Shank1}.

Our interest in this paper is to discuss the additional Umklapp
four fermion  coupling functions that arise when the lattice
vector ${\vec G}_i \neq 0$ in eq.(\ref{FSeq}). These satisfy
\begin{eqnarray}  \ \ {\vec P} \ &\equiv& \  {\vec K}_{1} \ + \
{\vec K}_{2} \nonumber \\ &=& \ \ {\vec P}_{i} \  \nonumber \\
&\equiv& {\vec K}_{3} \ + \  {\vec K}_{4} \ \pm G_{i}  \ \ \ , \
\  \ \   \ i \ = \ \pm 1 \ ,\pm 2 \ \ \ \end{eqnarray}
 with all
four ${\vec K}_{j}$ lying on the Fermi surface.

This requirement is most easily illustrated pictorially, as
shown in fig 2. Draw the DFS at $r \ = \ 2f(\theta)$, around the
origin O in momentum space. As argued above, any pair of initial
momenta on the Fermi surface will add up to a total momentum
${\vec P}$ which corresponds to some point P lying in the DFV
enclosed by this curve.  Now, draw four more images of this DFV
displaced respectively by the reciprocal lattice vectors $ G_{i}
$.  They will be centered respectively around $O_{1}, O_{2},
O_{3}$ and $O_{4}$, which are the images of O in the
neighbouring Brillouin zones. For sufficiently low density and
small $K_{F}(\theta)$, none of these image DFV will overlap with
the original DFV. At such low densities, as is well known, no
umklapp is possible.
 
But, for sufficiently high density such overlap will develop, as
depicted in fig. 2a. \ \big[ The threshold density for such
overlap to begin will depend on the details of the
single-particle energy $E({\vec K})$, but typically it should
begin to happen well before half-filling.\big]   When such
overlap exists, consider a total initial momentum ${\vec P} \ =
\ {\vec {OP}}$ which lies in the region of overlap between the
original DFV and one of its displaced images. Then the
correspondingly Brillouin displaced image of P, denoted by R in
fig.2a  will also lie inside the original DFV, and hence be an
acceptable total momentum for the final pair of states. Thus, an
umklapp process can happen in such a case. As illustrated in
fig.2b, if
\begin{equation} {\vec OP} \ \ \equiv {\vec P} \  \ = \ \ {\vec
K}_{1} \ + \ {\vec K}_{2 } \end{equation} 
then,
\begin{eqnarray} {\vec OR} \ \ \equiv {\vec P}_{1} \ \ &=& \ \
{\vec P} \ + \ {\vec G}_{1}   \nonumber \\ &=& \ \ {\vec K}_{3}
\ + \ {\vec K}_{4} \end{eqnarray} 
Note that in this umklapp process the final pair  ${\vec K}_{3}$
\ and \ ${\vec K}_{4}$ will be distinct from (although uniquely
determined by) the initial pair $ {\vec K}_{1}$ \ and \ ${\vec
K}_{2}$.  This in turn implies , for these momenta, an umklapp
coupling $F_{u}(\theta_{1}, \theta_{2})$ , which equals $
U({\vec K}_{1},{\vec K}_{2},{\vec K}_{3}, {\vec K}_{4})$ with
momenta as given in fig.2b. This is in addition to the usual
forward scattering coupling $F(\theta_{1}, \theta_{2})$ (which
formed the microscopic basis of the Fermi liquid parameter in
2-d).
  
	Such umklapp processes exist only for those values of
the initial total momentum ${\vec P}$ which lie in the region of
overlap between the original DFV and one of its Brillouin
displaced images. Such regions are shown dotted in the example
of fig.2a.  In the  shaded regions where the original DFV
overlaps with two different displaced images, each initial
${\vec P}$ will permit two different umklapp processes. In the
un-dotted, unshaded region in the interior of the original DFV
in fig.2a, only the forward coupling  is allowed except for
${\vec P} \ = \ 0$ where the so-called BCS coupling V is also
allowed. As the electron filling fraction increases the overlap
regions will increase. A stage will be reached when every point
in DFV, or equivalently, every pair of initial momenta on the
Fermi surface will yield one or more umklapp processes, in
addition to forward scattering. The threshold density at which
this begins to happen will depend on the precise shape of the
Fermi surface, but generically it should happen around
half-filling.
 
To illustrate this let us take the example of free fermions
half-filling a square lattice with nearest neighbour hopping
(see fig.3). The spectrum is
\begin{equation} E({\vec K}) \ = \ - cos(K_{x}) \ - \ cos(K_{y})
 \end{equation}
The reciprocal lattice vectors are $(0,\pm 2\pi)$ and $(\pm
2\pi, 0)$.  The Brillouin zone is a square of side $2\pi$ not
shown in the figure but its corners
$E=(-\pi.\pi),F=(\pi,\pi),G=(\pi,-\pi)$ and $H=(-\pi,-\pi)$ are
marked in fig.3.  The Fermi surface at half-filling is the
smaller tilted square shown in the figure whose vertices touch
the Brillouin zone .  The corresponding DFV is the larger tilted
square ABCD with vertices at $(\pm 2\pi \ , \ 0)$ and $(0 \ , \
\pm 2\pi)$ . Now, consider this DFV and its image shifted by one
of the reciprocal lattice vectors, say ${\vec G}_1 \equiv  \
(2\pi, 0)$ as shown by the dashed tilted square .  They overlap
in the  shaded region OFCG. For any generic point P in this
overlap region, its Brillouin shifted image $P_{1}$ also lies in
the original DFV and therfore a valid total momentum. The
initial pair $ {\vec K}_{1} \ + \ {\vec K}_{2}$ on the Fermi
surface whose total momentum is ${\vec OP}$  as well as the
distinct final pair $ {\vec K}_{3} \ + \ {\vec K}_{4}$ whose
total momentum is ${\vec {OP}}_{1}$ are also shown in fig.3. 
Thus any initial total momentum lying in OFGC will permit an
umklapp process, where momentum conservation is violated by the
reciprocal vector $(-2\pi, 0)$.  This overlap region OFGC
corresponding to the reciprocal vector $(-2\pi, 0)$ occupied
one-fourth (the right quadrant) of the original DFV.  By
similarly considering overlaps with the other 3 DFV images
shifted respectively by $(-2\pi, 0)$ and $(0,\pm 2\pi)$, one can
see that every initial momentum will lie in the overlap with one
or the other DFV image and can permit an umklapp final state.
Therefore, in this example, for
\underline{every}
 generic initial pair of paticles on the Fermi surface, there
will be an umklapp process in addition to the familiar forward
scattering coupling.  If the filling is somewhat less than half,
there will be a small region near the origin (corresponding to
${\vec P} \approx 0$ where the original DFV does not overlap
with any of its Brillouin shifted cousins. Barring this region,
Umklapp will happen for all other initial total momentum.  \big[
We chose this example because its geometry is simple. We are
aware it  carries other  problems like Van Hove singularities
and nesting vectors .  One can avoid these problems by choosing
slightly more than half filling, or anisotropic hopping but the
presentation will be more cumbersome. \big]

 In summary, we have given in this section a pictorial
discussion of when umklapp couplings exist. It is evident from
our discussion that for typical systems at a filling fraction of
about one-half, a substantial fraction of , if not all initial
states on the Fermi surface can yield umklapp processes in
addition to the  forward scattering process. Put together with
the tree level RG analysis of sec. 2, we see that these umklapp
couplings will remain marginal and can therefore
potentially play a role in the low energy properties of such
systems. Let us next extend their study to  the one-loop level.

\section{RG Flows at Higher Loops}

        This section is concerned with the RG analysis at one
loop level. The forward coupling $F(\theta_1, \theta_2)$ and
the BCS coupling $ V(\theta_1,
\theta_3)$ have been studied in ref.(1). We follow the same
procedure to study the role of $F_u(\theta_1, \theta_2)$. Since
the analysis is more transparent in the wave-vector space, we
present the case of isotropic Fermi surfaces in detail and
indicate the steps for generalisation to noncircular FS.

        The contribution of $F_u$ to $S_0$ is easily disposed
of, as it comes from the diagram (4a). Clearly only forward
coupling is allowed in this diagram, hence $F_u$ cannot
contribute to the renormalisation of any quadratic couplings.

        Next we consider the quartic couplings. We need study at
one-loop level only those couplings which survived at the
tree-level in our sec 2.  The one-loop diagrams that contribute
to these are shown in Figs.(4b) to (4d). Referring to these
diagrams note that all the internal lines have momenta in the
shell $S_{d\Lambda}$ and for the outer legs the magnitude of all
the momenta are set equal to $K_F$ and their frequencies to
zero.  Set $ {\vec Q} = {\vec K}_3 - {\vec K}_1, \ {\vec Q}' =
{\vec K}_4 - {\vec K}_1,$ and ${\vec P} = {\vec K}_1 + {\vec
K}_2$ (where we have used the three-vector notation to write
both wave-vector and frequency).  Note that the forward coupling
$F(\theta_1,\theta_2)$ is obtained in the limit $Q
\rightarrow  0$, the forward exchange coupling $F(\theta_2,
\theta_1) $is obtained in the limit $Q' \rightarrow  0$, the
 BCS coupling
$V(\theta_1, \theta_3)$ is obtained in the limit $P \rightarrow 
 0$, and
finally $F_u(\theta_1, \theta_2)$ is obtained when ${\vec K}_3 +
{\vec K}_4 = {\vec P} + {\vec G}_i$. Now the contribution of the
diagram (4b) to the quartic coupling is given by
\begin{eqnarray} dU &&\ ( {\vec K}_4, {\vec K}_3, {\vec K}_2, 
{\vec K}_1) \ = \
\int_{S_{d\Lambda}
} {d^2 K \over 2\pi} \int {d \omega \over 2\pi} \ \  
{U( {\vec K} - {\vec Q} , \  {\vec K}_3, {\vec K}, {\vec K}_1)
 \ U( {\vec K}_4, {\vec K}, {\vec K}_2, {\vec K}-{\vec Q})
  \over (i\omega - \epsilon({\vec K})) \  (i\omega - \epsilon 
  ({\vec K} - {\vec Q}))} \nonumber \\
  && + \ \int_{S_{d\Lambda}} {d^2 K \over 2\pi} \int {d \omega 
  \over 2\pi} \ \  
{U( {\vec K} - {\vec Q} +{\vec G} , \  {\vec K}_3, {\vec K}, 
{\vec K}_1)
 \ U( {\vec K}_4, {\vec K}, {\vec K}_2, {\vec K}-{\vec Q} +
  {\vec G})
  \over (i\omega - \epsilon({\vec K})) (i\omega - \epsilon 
  ({\vec K} - {\vec Q} + {\vec G}))}
\end{eqnarray}
where the first term corresponds to the normal process, and  the
second term corresponds to the Umklapp process. When ${\vec Q}
\rightarrow   0$, as required for the contribution to
$F(\theta_1, \theta_2)$, the integration over $\omega$ yields a
zero answer for the first term, as both the poles have the same
sign. For the second term the contributing phase space is shown
in Fig.(5) where the FS and its shell are drawn with origins at
$ {\vec 0}$ and at ${\vec R} = {-\vec G}$. The requirement that
the two poles have opposite signs restricts the phase space to
regions I, II, III and IV.  All these are of order ${1 \over
s^2}$ and hence make no contribution to the RG flow equation.
       
In order to consider the contribution of Eq.(4.1) to other the
couplings V and $F_u$, one notes that as long as ${\vec K}' -
{\vec K} \equiv {\vec R}$ with $R \neq 0$, the contributing
phase space is of order ${1 \over s^2}$ and thereby makes no
contribution to the RG flow equations. For V-coupling ${\vec P}
= 0$ and ${\vec Q}$ is generically nonzero which makes ${\vec R}
= \ {\vec Q}$ \  or  \ ${\vec Q} - {\vec G}$ for the first and
the second terms of eq.(4.1) respectively. It should also be
noted that for less than a certain filling fraction depending
upon the geometry of the FS the region around ${\vec P} = 0$ is
not subject to Umklapp scattering.  For $F_u$ coupling, for a
given ${\vec K}_1$ and ${\vec K}_2$, one one obtains ${\vec
K}_3$ and ${\vec K}_4$ using the construction of Fig.(2b). One
again sees barring an exception ${\vec Q} \neq 0$.  Thus this
diagram make no contribution to the flows of V and $F_u$. The
diagram (4c) is similar to (4b) except that ${\vec K}' = {\vec
K} - {\vec Q}'$. It is easily verified that contributions to F,
V, $F_u$ are all of order ${1 \over s^2}$
         
 Now we come to diagram (4d). Here ${\vec K}' = {\vec R} - {\vec
K}$ with ${\vec R} = {\vec P}$ or ${\vec P} + {\vec G}$. The
contribution of the diagram may be written as
\begin{eqnarray}  dU && \ ( {\vec K}_4, {\vec K}_3, {\vec K}_2,
 {\vec K}_1) \ = \
 {1 \over 2} \int_{S_{d\Lambda}} {d^2 K \over 2\pi} \int 
 {d \omega \over 2\pi} \ \  
{U( {\vec P} - {\vec K} , \  {\vec K}, {\vec K}_2, {\vec K}_1)
 \ U( {\vec K}_4, {\vec K}_3 , {\vec K}, {\vec P}-{\vec K})
  \over (i\omega - \epsilon({\vec K})) \  (-i\omega  - \epsilon 
  ({\vec P} - {\vec K}))} \nonumber \\
  && +  \  \ {1 \over 2} \int_{S_{d\Lambda}} {d^2 K \over 2\pi} 
  \int {d \omega \over 2\pi} \ \  
{U( {\vec P} - {\vec K} +{\vec G} , \  {\vec K}, {\vec K}_2, 
{\vec K}_1)
 \ U( {\vec K}_4, {\vec K}_3, {\vec K}_1, {\vec P}-{\vec K} 
 + {\vec G})
  \over (i\omega - \epsilon({\vec K})) (-i\omega - \epsilon 
  ({\vec P} - {\vec K} + {\vec G}))}
\end{eqnarray}
The frequency summation yields non-zero contribution only when
$\epsilon({\vec K})$ and $ \epsilon({\vec K}')$ have the same
sign. The scattering phase space is again determined in the same
fashion.  One replaces in fig (5) ${\vec K} - {\vec R}$ by its
negative and sets ${\vec R} = {\vec P}$ \ , or  \ ${\vec P} +
{\vec G} $.  Contributions come from four regions like $I'$
which are of order ${1 \over s^2}$. Now clearly for couplings F
and $F_u$, ${\vec P}$ is nonzero and one gets no contribution.
Finally consider the V-coupling for which ${\vec P} = 0$. Here
as discussed in Ref.(1), a finite contribution is obtained from
the first integral of Eq.(4.2), but the second integral does not
contribute. Thus the $F_u$ coupling does not affect the flow
equation for V either. To summarise, one find that upto
one-loop, couplings F and $F_u$ are marginally relevent, while V
is relevent and its flow is not affected by either F or $F_u$.

	Our discussion so far has been up to the one loop level.
We have  no rigorous results about higher loop diagrams or the
convergence of the loop expansion of the beta function, unlike
references 4 to 6. However, using the techniques of ${1\over N}$
expansion one can offer a heuristic argument, similar to the one
used in ref 1, to the effect that the umklapp coupling $F_u$
will continue to $\underline{not}$ contribute even in higher
loop diagrams to the beta functions of the forward, BCS, or
umklapp processes. Recall that in the RG limit, this theory can
be mapped into a  N flavour theory (with $ N \rightarrow \infty $)
by discretising  the angular integration into cells of width
\begin{equation} \Delta \theta  \ = \ {2 \Lambda \over K_{F}} \ = \ {2\pi
\over N} \end{equation} 
The interaction in this mapping becomes of order
${1 \over N}$.  Then, by counting powers of N, one can argue
(see ref. 1) that only certain sequences of "bubble" diagrams
will survive in the RG limit ( $\Lambda \rightarrow 0, N
\rightarrow \infty$).

	 The one-bubble diagram which we had just studied has
two vertices each bringing a power of ${1 \over N}$.  Then there
was integration over one internal momentum ${\vec K}$ as in eq.
4.1 or 4.2.  If the angle of this momentum can run freely over a
finite range as $ \Lambda \rightarrow 0$, that introduces a
power of N, giving the diagram a ${1 \over N}$ dependence
altogether, of the same order as the tree diagram. This can
happen only if the vector ${\vec R}$ of fig. 5 is zero.  On the
other hand , when ${\vec R}$   is not zero, i.e.  when the
internal momentum angles are constrained by the external
momenta,  they do not bring another power of N. Thus, for the
forward coupling F the sequence that survives is a product of
bubbles with purely F vertices, while for the BCS couplin V, it
is a product of bubbles with purely V vertices. Whereas,
whenever a $F_u$ vertex is present, the vector ${\vec R}$ is
non-zero and the diagram is down by a power of ${1 \over N}$.

	Hence, the umklapp coupling $F_u$ does not contribute at
any loop order to the beta function of any of the couplings ,
including itself. However, at the tree level, we saw that it is
marginal and survives. Admittedly, these arguments are only
heuristic. We have not examined questions of convergence of the
perturbation series unlike references 4 to 6.     

       From these arguments the generalisation to noncircular
Fermi surfaces possessing the time-reversal symmetry is quite
straight forward. The contribution in this case also depends on
the value of the momenta ${\vec R}$. Whenever ${\vec R}$
determined by external legs is nonzero the scattering phase
space is again of order ${1 \over s^2}$ .It is easily verified
that the above conclusion holds for such surfaces too.

\section{Conclusions}
 	We have developed above a method for determining when
and to what extent umklapp couplings will survive as marginally
relevant interactions for electrons in 2 dimensional one-band
systems on a lattice.  We found that for densities of the order
of one-half per site, most (if not all) initial pairs of
particles near the Fermi surface are kinematically permitted to
undergo umklapp scattering. We also showed that under RG flow
these umklapp couplings , if present in the original starting
Hamiltonian , will remain marginal , along with
forward scattering and the BCS coupling. Therefore , unless
neglected for some reason in the original microscopic
Hamiltonian, umklapp couplings will very much be copiously
present in the effective low energy Hamiltonian obtained in the
$s \rightarrow \infty$ limit of the RG flow, for densities near
one-half.

	 These results were obtained in 2 dimensions, but as
mentioned in  sec.III, they can be generlised to 3 dimensions in a
straightforward way. Given  a pair of particles on a convex
time-reversal invariant Fermi surface $r=f(\theta, \phi)$, the
domain of their sum ${\vec P}$ will be the DFV enclosed by
$r=2f(\theta, \phi)$ . The overlap of this DFV with its partners
shifted by various reciprocal lattice vectors will again give
${\vec P}$-domains of umklapp processes. Whereas in 2D,
generically there will be a unique pair of final momentum
directions for umklapp, given an initial pair, in 3D there will
be a whole cone of allowed final directions for umklapp, because
the intersection of two Fermi surfaces will be closed curve
rather than a pair  of points.  This azimuthal degree of freedom
for the final states will be available for umklapp for the same
reason as happened for normal (non-umklapp) couplings
\cite{Shank1},

	Given that umklapp couplings will survive in the low
energy effective Hamiltonian, it is bound to affect some
physical properties of such systems such as resistivity, thermal
conductivity etc.  For instance, in the case of one band
Hamiltonians such as ours, electron-electron interaction cannot
lead to resistance unless some umklapp process takes place. The
well known $T^2$ dependence of resistance can arise for such
systems only because of umklapp.  As far as the Landau Fermi
liquid theory is concerned, its existing formulation seems to involve
only the forward scattering amplitude. In that theory, low
energy excitations of the system are described in terms of
quasi-particles with a weak residual interaction  of the form
$f({\vec p}, {\vec p}') \delta n_{\vec p} \delta n_{{\vec p}'}$,
where $\delta n_{\vec p}$ is the change in the occupation of the
${\vec p}$-state from the ground state.  This form of the
interaction  involves only the occupation numbers and hence only
the forward scattering amplitude.  Indeed in the
detailed many-body perturbative analysis $f({\vec p}, {\vec
p}')$ has been identified with the scattering amplitude
$T(p+q,p'-q,p,p')$ under the zero momentum transfer limit
$({\vec q}, \omega) \rightarrow   0$ such that ${q \over \omega}
\rightarrow 0$ . Now, as shown in sec.IV , the higher order contributions of
 $F_{u}$ to the forward coupling F vanishes in the $s 
\rightarrow \infty$ limit of the RG flow. Hence the Landau interaction
$f({\vec p}, {\vec p}')$ will not , in and of itself, be affected by
the presence of umklapp. However, the interaction $F_u$ is marginal.
 It will have to be included by modifying the Fermi liquid Hamiltonian.

	Similarly, whenever there is only a soft response
 $R\big(Q\rightarrow 0,
\omega \rightarrow 0 \big)$ to a soft probe, only the forward
coupling matters.  
But, in a lattice, as compared to a spatial continuum,
a soft probe $J \big(Q\rightarrow 0, \omega \rightarrow 0)
\big)$ can lead to a response $R\big(Q =G , \omega \rightarrow
0) \big)$ in the presence of umklapp. In such cases, the fact
that umklapp coupling remains marginal will affect
the physics.

\begin{figure}
\caption{Construction of the Double Fermi Volume (DFV). The DFV
is the ${\vec K}$-space region allowed to vector ${\vec P} =
{\vec K}_{1} + {\vec K}_{2}$, where ${\vec K}_{1}$ and ${\vec
K}_{2}$ are any pair of vectors lying on the Fermi surface. The
FS, given in polar coordinates by $r = f(\theta)$ is denoted by
the solid curve.  DFV is the region inside the dotted curve $r =
2 f(\theta)$. Fig.(a) supports the argument that all vectors
${\vec P}$ must lie in DFV provided FS is a convex surface.
Fig.(b) illustrates the converse argument that any point P
inside the DFV can be sum of two vectors on the FS.}
\label{autonum}
\end{figure} 

\begin{figure}
\caption{Regions of total momentum ${\vec P}$ (of two particles)
for which Umklapp processes are allowed. The diagram (a) shows
DFV and its four images, which are DFV's drawn with their
origins shifted by four reciprocal vectors of the square
lattice. The lines (not drawn) $OO_{1}$, $OO_{2}$, $OO_{3}$, and
$OO_{4}$ denote the four reciprocal vectors.  The dotted regions
for which DFV and any of its images overlap are those values of
${\vec P}$ for which an Umklapp process is allowed.  For the
shaded regions the DFV overlaps with two of its images, and for
these two Umklapp processes are permitted. The diagram (b) shows
the construction of outgoing momenta ${\vec K}_{3}$ and ${\vec K
}_{4}$, when an Umklapp process occurs. Also shown are the
incoming momenta ${\vec K}_{1}$ and ${\vec K}_{2}$ and their sum
${\vec P}$.}
\end{figure}

\begin{figure}
\caption{Fermi surface (inner diamond) and DFV (shown as ABCD)
for a half-filled square lattice with $E({\vec K})$ given by
Eq.(3.6). The Brillouin zone is the square whose vertices are
EFGH.  The dashed square is one of the images of DFV and the
shaded region is the P-region for which Umklapp by vector
(-2$\pi$,0) is allowed. Clearly the other 3 images will overlap
with the remaining portions of DFV, thus permitting Umklapp for
every point of DFV. Also shown are ${\vec K}_{3}$ and ${\vec
K}_{4}$ for a typical set of ${\vec K}_{1}$ and ${\vec K}_{2}$.}
\end{figure}

\begin{figure}
\caption{The diagrams contributing to couplings at one loop
level.  The internal lines carry momenta from the shells
$S_{d\Lambda}$. Fig.(a) gives the contribution to two-point
couplings, while (b), (c) and (d) denote the three diagrams
contributing to four-point couplings.}
\end{figure}

\begin{figure}
\caption{This diagram gives  the momentum phase space regions
that are allowed for processes depicted in diagrams from
Fig.(4b) to (4d). For Fig.(4b) the vector ${\vec R}$ takes
values ${\vec Q}$ and ${\vec Q} + {\vec G}$. For Fig.(4c) ${\vec
R} = {\vec Q}' $ and ${\vec Q}' + {\vec G}$. The allowed values
of ${\vec K}$ are from regions denoted by I, II, III, and IV.
For Fig.(4d) ${\vec R} = {\vec P}$ and ${\vec P} + {\vec G}$ and
the sign of ${\vec K} - {\vec R}$ is to be reversed. The allowed
values of ${\vec K}$ come from four regions like $I'$.}
\end{figure}

\end{document}